# Anomalous Hall and Nernst Effects in FeRh

Hilal Saglam[1, 2, *, †], Changjiang Liu[1, *], Yi Li[1], Joseph Sklenar[3], Jonathan Gibbons[4], Deshun Hong[1], Vedat Karakas[1, 5], John E. Pearson[1], Ozhan Ozatay[1, 5], Wei Zhang[6], Anand Bhattacharya[1], and Axel Hoffmann[1, 4]

[1]Materials Science Division, Argonne National Laboratory, Lemont, Illinois 60439, USA
[2]Department of Applied Physics, Yale University, New Haven, Connecticut 06511, USA.
[3]Department of Physics and Astronomy, Wayne State University, Detroit, Michigan 48202, USA
[4]Materials Science and Engineering, University of Illinois at Urbana-Champaign, Urbana, Illinois 601801, USA
[5]Department of Physics, Bogazici University, Istanbul 34342, Turkey
[6]Department of Physics, Oakland University, Rochester, Michigan 48309, USA

\* Denotes equal contribution
† Correspondence to hilal.saglam@yale.edu

Antiferromagnets with tunable phase transitions are promising for future spintronics applications. We investigated spin-dependent transport properties of FeRh thin films, which show a temperature driven antiferromagnetic-to-ferromagnetic phase transition. Epitaxial FeRh films grown on MgO (001) substrates exhibit a clear magnetic and electronic phase transition. By performing anomalous Hall and anomalous Nernst effect measurements over a wide range of temperatures, we demonstrate that the thermally driven transition shows distinctly different transverse transport on both side of the phase transition. Particularly, a sign change of both anomalous Hall and Nernst signals is observed.

Recently, antiferromagnets have attracted increasing attention due to their promising properties, especially for spintronics devices[1,2,3]. For instance, the spin precession in antiferromagnets typically occurs in the elusive THz frequency regime, which is crucial for high-speed operations. Also, antiferromagnetic order is robust against external magnetic fields due to the vanishing net magnetization. Conversely, the absence of a net magnetization implies automatically that it is non-trivial to control antiferromagnets simply via external fields as in the case of ferromagnets. Towards this end, antiferromagnets have attracted tremendous interest due to their anomalous optical and transport phenomena[4]. For instance, despite their vanishing magnetization, noncolinear antiferromagnets, e.g., hexagonal $Mn_3Ge$ and $Mn_3Sn$ show large anomalous Hall effect (AHE)[5,6], anomalous Nernst effect (ANE)[7], and magneto-optical Kerr effect[8]. Furthermore, the magnetic spin Hall effect has been observed in hexagonal $Mn_3Sn$[9] and cubic $Mn_3Ir$[10], which both have non-collinear spin structures. In addition, it has recently been shown that the topological states in $Mn_3Sn$ can be electrically manipulated[11]. Lastly, similar anomalous Hall and magneto-optic Kerr effects are also expected for certain collinear antiferromagnets, which is a promising pathway for electrical detection and manipulation of antiferromagnetic spin structure[12,13,14].

Among various antiferromagnets, equiatomic FeRh is of particular interest and holds promises for novel spintronics applications due to its intriguing properties[15]. The B2-ordered FeRh undergoes a temperature driven antiferromagnetic-to-ferromagnetic phase transition [see Fig. 1(a)]. This transition is accompanied by a pronounce resistivity change[16], an isotropic volume expansion,[17] as well as a large entropy release[18]. Besides, the transition temperature can be tuned via external sources such as chemical substitutions[19,20], magnetic field[21], lattice strain[22], and pressure[23]. By performing Hall and thermoelectric power measurements in FeRh[24] and Fe(3.5%Ni)Rh[20], respectively, it has been shown that both Hall coefficient and thermoelectric power change their sign through the phase transition. Due to these advantages, this material has gained interest for heat assisted information storage[25] and magnetocaloric cooling applications[26]. Nowadays, FeRh attracts renewed interest for current induced spintronics applications due to its large spin orbit coupling arising from Rh[27,28]. Towards this end, recent experiment shows that the Néel order of FeRh can be controlled via a specific field-cooling protocol, which yields two distinctive resistance states[29]. Other experimental work suggests a current induced phase transition, where sequential write-read operation using only electric current and external field has been demonstrated. Here, the writing mechanism relies on the antiferromagnetic-ferromagnetic transition induced by Joule heating[30]. More recently, by performing spin-pumping and inverse spin Hall effect experiments in FeRh/NiFe and FeRh/Pt bilayers, it has been shown that FeRh can act both as a spin sink and spin detector, respectively[28]. In addition, first-principle theory predicts a sign change of the spin Hall conductivity of FeRh across the transition. However, the same work suggests a suppression of anomalous Hall and Nernst conductivities in antiferromagnetic FeRh[27].

Although the metamagnetic transition in FeRh was first reported over a half century ago[31], the origin of it is still an ongoing debate. There are various mechanisms reported in the literature, such as a change in the electronic density of states at the Fermi level[32,33], spin wave excitations[34] and Rh instability[35]. By hard X-ray photoemission spectroscopy, the argument of the increase in density of states at the Fermi level in the FM phase has been observed experimentally[36], which is also supported by specific heat measurements[37]. On the other hand, surface sensitive soft X-ray measurements reveal a very little modification in the electronic structure of the FeRh[38]. In this work, we have investigated the anomalous Hall and anomalous Nernst effects in FeRh at various temperatures that cover its magnetic phase transition. As will be discussed in detail below, this phase transition gives rise to significant modifications of the FeRh electronic structure. In particular, we observe a sign change of both anomalous Hall and Nernst signals when FeRh switches from the antiferromagnetic to the ferromagnetic state.

We have grown epitaxial FeRh thin films on MgO (001) using magnetron sputtering[39]. In order to desorb any contamination on the MgO surface, the single crystal substrates were annealed at 825 °C for one hour prior to the deposition. Then the temperature was decreased to 450 °C for another hour in order to stabilize the deposition temperature. Afterwards, we introduced 65 sccm Ar gas into the deposition chamber and kept the deposition pressure at 6 mTorr during the growth. Applying 50-W of magnetron power yielded an overall growth rate of 0.7 Å/s. After the deposition was completed, we increased the temperature to 650 °C for another hour in order to post anneal our films to improve their crystalline quality.

The measured X-ray diffraction (XRD) pattern of a 20-nm thick FeRh is shown in Figure 1(b). The pronounced (001) super-lattice and (002) fundamental peaks for the FeRh indicate the formation of chemically ordered B2 structure[40]. From the Bragg peak positions, we extracted the lattice constants of MgO and FeRh as 0.421 nm and 0.30 nm, respectively. This confirms that the lattice mismatch between MgO and FeRh is relatively small around 0.3 %, if one considers that the unit cell of the FeRh is rotated 45° in-plane relative to the unit cell of the MgO[41]. Figure 1(c) shows measured and fitted[42] X-ray reflectivity (XRR) spectra of a continuous FeRh film as a function of incident angle, from which the thickness, roughness and density of the film were extracted as 20.07±1.13 nm, 0.34±0.08 nm and 0.035±0.001 atoms/Å$^3$, respectively.

SQUID magnetometry was used to investigate the meta-magnetic phase transition as a function of temperature. Figure 1(d) shows the first order phase transition for a 20-nm thick continuous FeRh film. In this particular experiment, we swept the temperature from 395 K to 150 K and vice versa in the presence of a 0.2-T magnetic field. Note that a residual ferromagnetic phase is observed in the nominal antiferromagnetic state, which has been reported previously for thin FeRh films[43]. One explanation for this remnant magnetization is that a residual, chemically disordered phase of FeRh is possibly distributed throughout the entire film, ensuring that some ferromagnetism persists below the transition temperature[19]. Another possibility is that a ferromagnetic layer is situated at the interface between the MgO and FeRh layer, where the chemical ordering and/or strain of the film differs from the rest of the film and this likely stabilize a ferromagnetic phase below the ferromagnetic-to-antiferromagnetic transition temperature[44]. The metamagnetic phase transition is also confirmed by four terminal resistance measurements as seen in Figure 1(e), where we observe an abrupt increase in resistance as the FeRh goes through the ferromagnetic to antiferromagnetic ordering transition.

In order to perform anomalous Hall effect (AHE) and anomalous Nernst effect (ANE) measurements, we patterned our FeRh films into Hall bar devices using photolithography and subsequent ion milling. Figure 5(a) show the anomalous Hall and Nernst effect measurement geometries for a 20-nm thick FeRh film with lateral dimensions of 10×900 μm$^2$. For AHE measurements, a 10-μA current was applied using a current source and the concomitant transverse voltage was measured by a nano-voltmeter. During the Hall (anomalous) measurements, an out of plane external field was swept. In order to correct the misalignment due to the photolithography, the measured transverse voltages were antisymmetrized with respect to magnetic field to eliminate the symmetric component from magnetoresistance. For anomalous Nernst effect measurements, we used an on-chip heating technique[45], where a 50-nm Au layer was used as a heater. In order to ensure electrical insulation between the Au heater and FeRh film, a 150-nm MgO layer was grown between the Au and FeRh. The local heating was achieved by applying a voltage of 0.53 V$_{rms}$ at 3 Hz to the Au heater layer over a 65 Ω load resistor, which yields a heater power of approximately 1 mW. Simultaneously, a magnetic field was swept in the plane of the sample and the resultant second harmonics component of the voltage is measured using a lock-in amplifier at the second harmonic frequency (6 Hz). Note that the resulting temperature gradient is in the out of plane direction and the temperature gradient, magnetic field and the resultant electric field are all perpendicular to each other.

**ANOMALOUS HALL EFFECT**

We first discuss the magnetoresistance (MR) behavior of FeRh across its phase transition, which is shown in Figures 2(a-c) for 100, 270 and 390 K, respectively. Here, MR is measured as an in plane external field is swept perpendicular to the current. Starting with Figure 2(a), we notice that FeRh exhibits a negative MR at low fields, which is inconsistent with previous experiments[46], where the MR in the AFM phase was reported as solely positive. This low field behavior in the antiferromagnetic state may suggest that the external magnetic field suppresses the spin disorder scattering leading to a negative MR by orienting the residual magnetization along the field direction. On the other hand, the high field MR in the AFM state reveals approximately quadratic positive behavior, which can be associated with the ordinary magnetoresistance that is spin-independent[29]. In Figure 2(b), we show the MR at the temperature that FeRh is in the mixed state. This hysteresis behavior in the mixed state was also observed in a recent study and suggested that with increasing field FeRh spins approach the ferromagnetic coupling[29]. Figure 2(c) shows the high temperature behavior (ferromagnetic state) in which the negative MR increases abruptly at low magnetic fields but varies smoothly at higher fields. The non-saturating MR at higher fields is likely due to a magnetic field-induced suppression of magnon excitations, consistent with previously reported results for other ferromagnetic metals[47]. On the other hand, the low field magnetoresistance is coming from the alignment of magnetization, i.e., conventional anisotropic magnetoresistance effect.

Now, we discuss the Hall effect measurements in FeRh. Figure 3(a) shows measured transverse voltages as a function of magnetic field at various temperatures. We observe that the measured voltages increase linearly at low magnetic fields in both ferromagnetic and antiferromagnetic states, which is consistent with a hard-axis magnetization response due to shape anisotropy and then crosses over to a constant region at high fields. The change in sign and magnitude of the slope of the Hall voltage at high fields suggest that the majority charge carrier type and density alter over the phase transition, consistent with previously reported results[24,48]. In order to quantify this, we plotted the Hall coefficient at different temperatures [see Fig. 3(b)]. It is remarkable that the magnitude of the Hall coefficient increases by roughly a factor of three as the FeRh transits from the ferromagnetic to the antiferromagnetic state, which implies a large change in the electronic structure of FeRh across the phase transition.

In order to reveal the anomalous Hall effect contribution to the measured total voltages, we subtracted the linear backgrounds at high fields, which is due the ordinary Hall effect as discussed above [see Fig. 4(a)]. We observe a non-vanishing anomalous Hall signal in the antiferromagnetic phase, and more interestingly the signal reveres its sign as FeRh becomes an antiferromagnet, in agreement with a recent work[48]. However, the same work reports a hump-like anomaly in their Hall measurements in the antiferromagnetic state and attributes this to the topological Hall effect (THE). Note that we have not observed explicitly such behavior in our measurements [see Fig. 3(a)]. The non-vanishing anomalous Hall voltage at low temperatures in the AFM state can potentially be associated with the emergence of a non-collinear spin texture due to the exchange coupling between the antiferromagnetic order and the residual magnetization. This non-vanishing anomalous Hall signal was further corroborated by the coercivity enhancement in the antiferromagnetic state. Figure 4(b) shows the extracted coercivity as a function of temperature, where we observe an almost zero coercivity in the ferromagnetic phase, while there is a pronounced enhancement at low temperatures at which FeRh is an antiferromagnet. This enhancement in coercivity can be linked to the exchange interaction between the residual magnetization and the bulk antiferromagnetic order at low temperatures, which has also been observed recently in another study[49].

Next, we plot the anomalous Hall angle [see Fig. 4(c)], which provides a quantitative determination of the strength of the anomalous Hall effect and is described by the ratio of $\sigma_{xy}$ and $\sigma_{xx}$. Here, $\sigma_{xy}$ and $\sigma_{xx}$ are described by

$$\sigma_{xx} = \frac{\rho_{xx}}{\rho_{xx}^2 + \rho_{xy}^2}, \qquad (1)$$

$$\sigma_{xy} = -\frac{\rho_{xy}}{\rho_{xx}^2 + \rho_{xy}^2}, \qquad (2)$$

where, $\sigma_{xx}$ and $\sigma_{xy}$ are the longitudinal charge conductivity and anomalous Hall conductivity, respectively. And $\rho_{xx}$ and $\rho_{xy}$ are the longitudinal and transverse resistivities, respectively. The temperature dependent behavior of the anomalous Hall angle has a negative sign at low temperatures, shows a linear increase in the mixed-state and then tends to become constant at high temperatures.

The contributions to the anomalous Hall signal can be divided broadly into two categories, i.e., intrinsic effects such as the Berry phase and extrinsic effects such as skew scattering and side jumps[50,51]. These contributions dominate in different conductivity regimes and can be categorized by a specific power-law relation between the anomalous Hall conductivity and the longitudinal charge conductivity. For high-conductivity metals, the major contribution comes from the skew scattering, while in alloys or disordered materials with moderate and low conductivity, the intrinsic or side-jump mechanisms dominate, respectively[50,51]. In order to better understand the non-vanishing anomalous Hall signal in FeRh, we analyzed our data using this scalability relation. Figure 4(d) shows $\sigma_{xy}$ as a function of $\sigma_{xx}$ for different temperatures. It is seen that $\sigma_{xy}$ is nearly independent of $\sigma_{xx}$ at temperatures below 250 K, where the FeRh is in its antiferromagnetic state. Then it scales linearly in the mix-state and becomes almost constant again at higher temperatures (in ferromagnetic state). This independence of $\sigma_{xy}$ with respect to both temperature and $\sigma_{xx}$ might suggest that the sign change observed originates from intrinsic effects, i.e., scattering independent mechanisms.

**ANOMALOUS NERNST EFFECT**

The thermoelectric counterpart of the anomalous Hall effect is known as the anomalous Nernst effect (ANE), which is expected to scale with the magnetization of the material. For ordinary ferromagnetic material, NE and ANE effect are typically described by the following empirical expressions[52];

$$\vec{E}_{NE} = Q_0(\vec{B}_{ext} \times \nabla T) \tag{3}$$

$$\vec{E}_{ANE} = Q_s(\mu_0 \vec{M} \times \nabla T) \tag{4}$$

where, $Q_0$, $\vec{B}_{ext}$, $\nabla T$ and $\vec{E}_{NE}$ are the Nernst coefficient, the external magnetic field, the thermal gradient across the ferromagnet, and the Nernst effect induced electric field, respectively while $Q_s$, $\mu_0$, $\vec{M}$ are the anomalous Nernst coefficient, vacuum permeability and the magnetization of the ferromagnet, respectively.

Figure 5(b) shows the measured anomalous Nernst voltages as a function of external magnetic field for different temperatures. At high temperatures, we observe steep jumps at low fields and almost constant curves at higher fields, consistent with the typical behavior expected from ferromagnets. As the temperature lowers and the FeRh becomes an antiferromagnet, one would expect to observe a vanishing ANE signal due to suppression of the spontaneous magnetization. However, we measure non-zero voltages at lower temperatures and more interestingly the signal reverses its sign compared to the ferromagnetic state. Note that both the anomalous Hall and Nernst signals show nearly the same field dependence, which are similar to the magnetization curve, implying that the non-vanishing signals are associated with the residual magnetization in the antiferromagnetic state. Also, note that the magnetic field dependence of the anomalous Nernst effect has a different behavior compared to the anomalous Hall effect. This is because the external field was applied in the plane of the sample during the anomalous Nernst measurements.

We also evaluated the anomalous Nernst coefficient of the FeRh across its magnetic phase transition. To do so, we first determined the heat gradient across the sample using the differential form of Fourier's Law, which connects the heat flux density, $\Phi$ to the temperature gradient via the thermal conductivity, $K$.

$$\Phi = -K\frac{dT}{dz}, \tag{5}$$

where, the temperature-dependent thermal conductivity for FeRh thin film was extracted using the Wiedemann-Franz law,

$$\frac{K}{\sigma} = LT, \tag{6}$$

where, $K$ is the thermal conductivity, $\sigma$ is the electrical conductivity, $T$ is the temperature and $L$ is the Lorentz number. Figure 5(d) shows the extracted Nernst coefficient across the phase transition of the FeRh. As also observed in the anomalous Hall effect measurements, the sign of $Q_s$ is negative at lower temperatures and becomes positive once the FeRh enters the mixed state. Subsequently, it increases smoothly as the FeRh becomes more ferromagnetic. The magnitude of the anomalous Nernst coefficient in the FM phase is comparable to the conventional ferromagnetic metals[53].

In conclusion, we investigated anomalous Hall and Nernst effects in epitaxially grown FeRh films. Our experimental results show that the thermally driven transition leads to significant modifications to the electronic transport properties of FeRh across the transition temperature. Specifically, we observe that the charge carrier density and type change significantly across the transition. Besides, a sign change of both anomalous Hall and Nernst signals is observed. The sign change of anomalous Hall and Nernst signal not only opens up a different pathway for the further exploration of the nontrivial spin structures but also may pave the way for potential applications in AFM spintronics.

**ACKNOWLEDGMENTS**

All experimental work was supported by the U.S. Department of Energy, Office of Science, Basic Energy Sciences, Materials Science and Engineering Division. Lithographic patterning was carried out at the Center for Nanoscale Materials, which is supported by the DOE, Office of Science, Basic Energy Sciences under Contract No. DE-AC0206CH11357. Part of the manuscript preparation was supported as part of Quantum Materials for Energy Efficient Neuromorphic Computing, an Energy Frontier Research Center by the U.S. DOE, Office of Science and by the NSF through the University of Illinois at Urbana Champaign Materials Research Science and Engineering Center DMR-1720633. Vedat Karakas and Ozhan Ozatay acknowledge financial support from TUBITAK grant 118F116 and Bogazici University Research Fund grant 17B03D3. Hilal Saglam and Axel Hoffman thank to Sheena Patel and Eric Fullerton for useful discussions about FeRh growth. Hilal Saglam thanks to Carlo Segre and Kamil Kucuk for their help with the XRD analyses.

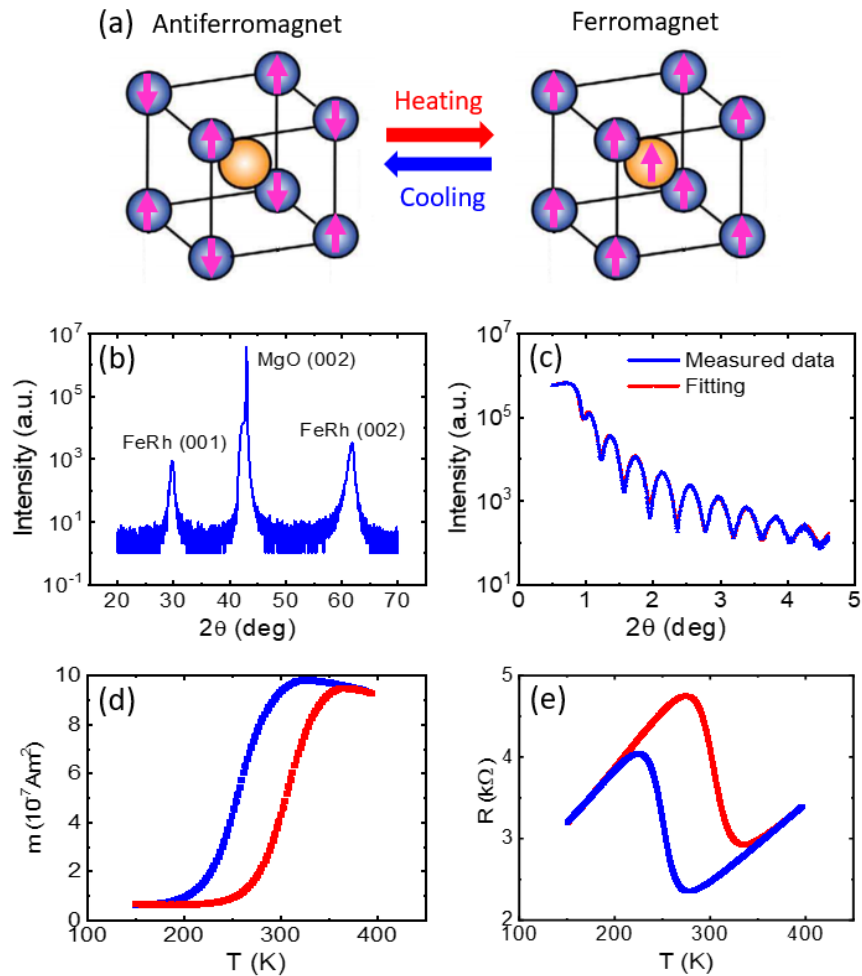

**FIG. 1**. (a) Antiferromagnetic and ferromagnetic FeRh unit cells in which the dark blue spheres symbolize Fe atoms and yellow spheres represent Rh atoms with the direction of magnetic moments indicated by pink arrows. (b) X-ray diffraction pattern of a 20-nm thick FeRh grown on a (100) MgO substrate. The pronounced peak for (001) FeRh confirms the desired B2 ordering. (c) X-ray reflectometry and simulation data from which the thickness of the film is extracted as approximately 20 nm. (d) Net magnetic moment versus temperature for a 20-nm thick continuous FeRh film measured in a presence of a 0.2-T magnetic field. Red and blue lines represent warming and cooling cycles, respectively. (e) Change in resistance as a function of temperature for a 20-nm FeRh film patterned into a Hall bar structure.

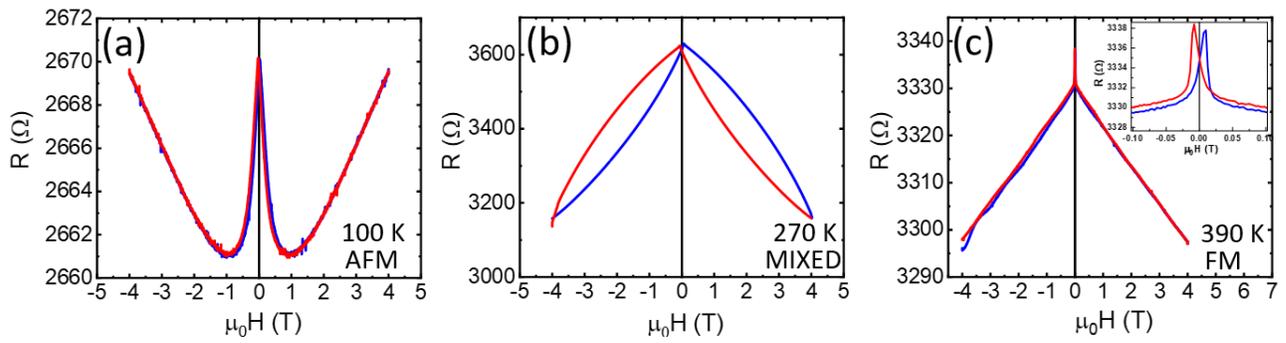

**FIG. 2**. In plane magnetoresistance behavior of FeRh measured at different temperatures: (a) 100 K, (b) 270 K, and (c) 390 K. The magnetic field was applied perpendicular to the current direction. The blue and red lines represent the field sweeping from negative to positive and positive to negative fields, respectively.

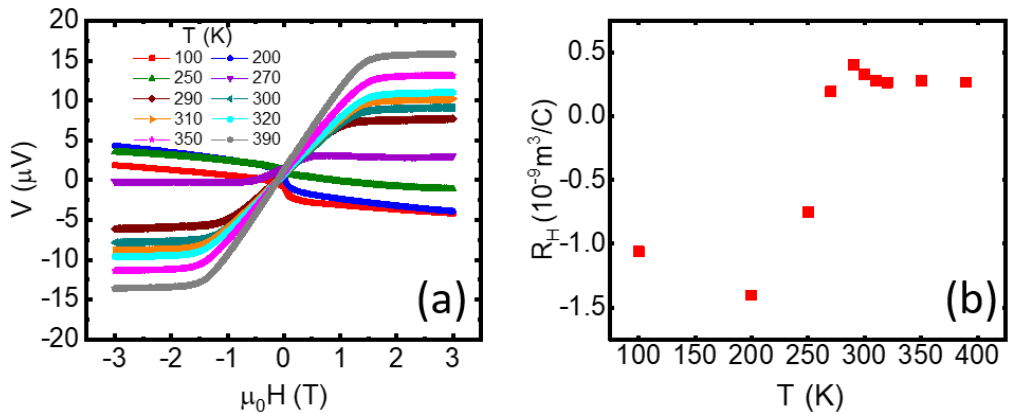

**FIG. 3.** (a) Measured transverse voltages as a function of magnetic field for different temperatures. (b) Extracted Hall coefficient at various temperatures.

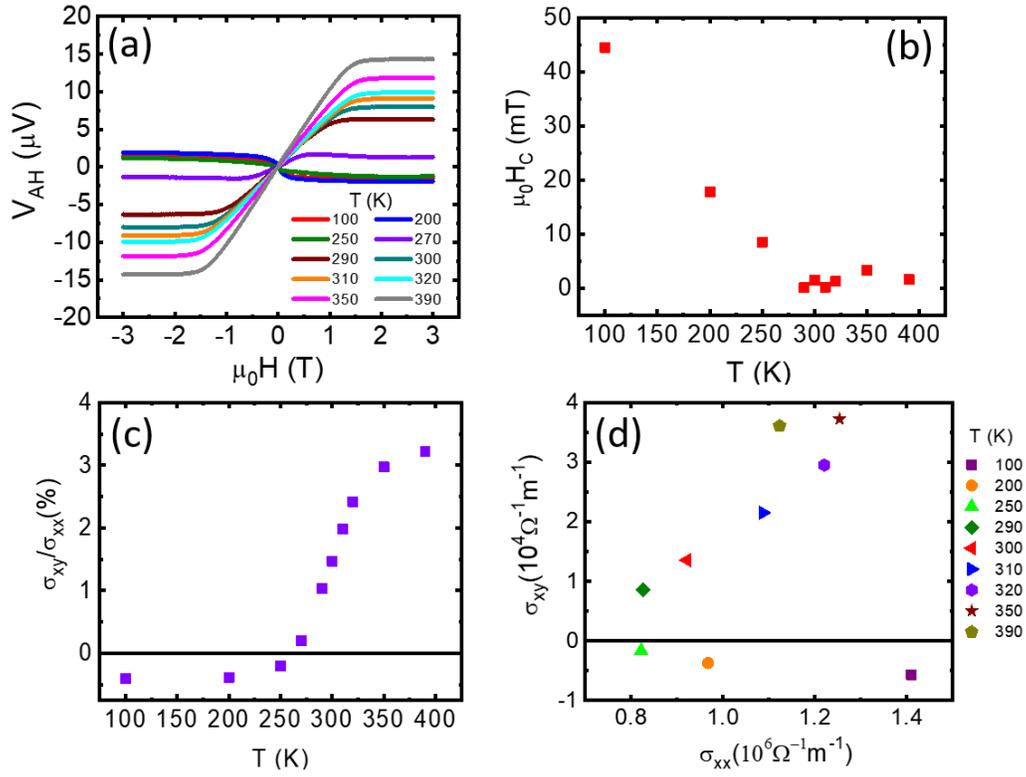

**FIG. 4.** (a) Measured anomalous Hall voltages as a function of magnetic field for different temperatures obtained by subtracting the high field linear dependence from the data in Fig. 3(a). (b) The change in coercivity as a function of temperature extracted from anomalous Hall voltages. (c) Anomalous Hall angle for varying temperatures that cover the metamagnetic transition. (d) Temperature dependence of the anomalous Hall conductivity as a function of the conductivity.

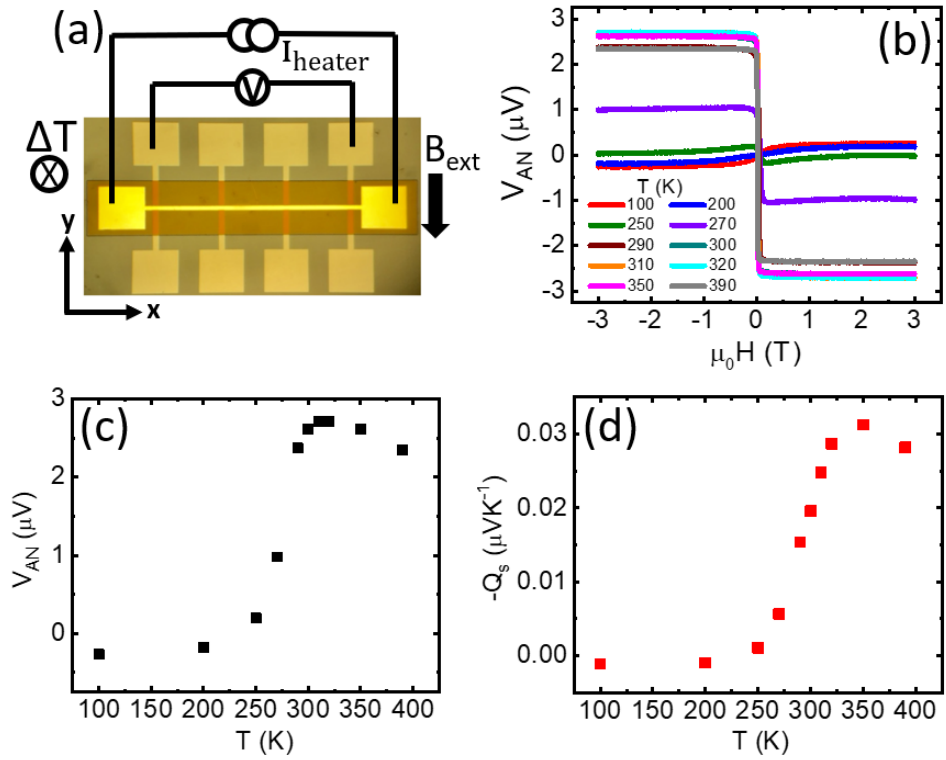

**FIG. 5.** (a) Schematic of the Nernst measurement setup. (b) Measured anomalous Nernst effect as a function of magnetic field for different temperatures. (c) Magnitude of the anomalous Nernst voltages for varying temperatures. (d) The change in anomalous Nernst coefficient as a function of temperature extracted from the anomalous Nernst voltages.